\documentclass[11pt,a4paper]{article}
\usepackage{amsmath,amssymb,amsfonts,a4wide,graphicx,bm,times,psfrag,wrapfig,sidecap,
adjustbox, yfonts} \usepackage{cite}
\usepackage[colorlinks=true,linkcolor=black, citecolor=black,
urlcolor=black]{hyperref} \numberwithin{equation}{section}
\makeatletter \let\old@startsection=\@startsection
\renewcommand{\@startsection}[6]
{\old@startsection{#1}{#2}{#3}{#4}{#5}{#6\mathversion{bold}}}
  
\def\defeq{\stackrel{\text{def}}=} \newcommand\re[1]{({\ref{#1}})}
\def\be{\begin{eqnarray} } \def\ee{\end{eqnarray}}
\def\IK{{\mathbb{K}}} \def\ID{{\mathbb{D}}} \def\no{\nonumber}
\def\la{\label} \def\({\left(} \def\){\right)} \def\<{\langle}
\def\>{\rangle} \def\[{\left[} \def\]{\right]} \def\p{\partial}
  \def\g{\gamma} 
   \def\vp{\varphi}
  
\def\IR{{\mathbb{R}}}\def\IZ{{\mathbb{Z}}} 
 \def\CA{{\cal A}} 
\def\CB{{\cal B}} \def\CO{{ \mathcal{ O} }} \def\CZ{{ \mathcal{ Z} }}
  
 \def\CH{ {\cal H}}   \def\CF{{\cal F}}  \def\Tr{{\rm Tr}}  \def\d{\delta} \def\uu{ {\bf u}} \def\ww{ { \bf w }}
\def\vv{ {\bf v}} \def\bA{\mathbf{A}} \def\bB{\mathbf{B}}
 \def\rr{ \mathbf{r}}
 \def\bH{\mathbf{H}} \def\bO{ \mathbf{\Omega} }
\def\bvp{{\bm{\vp}}}
\def\bpsi{\bm{\psi}}\def\bbvp{ {\bm{\bar \vp}}}\def\bbpsi{\bm{\bar
\psi}} \def\redr{ {\color{red}\rho} }

\def\LR{ { \!_{L,R}}} \def\ant{ {\dot{ : }}}
                
\usepackage{blindtext}
 
\usepackage{xcolor}
\colorlet{red}{red!55!black}
\colorlet{blue}{blue!90!black}
 
\newcommand\encadremath[1]{\vbox{\hrule\hbox{\vrule\kern8pt
\vbox{\kern8pt \hbox{$\displaystyle #1$}\kern8pt}
\kern8pt\vrule}\hrule}} \def\enca#1{\vbox{\hrule\hbox{
\vrule\kern8pt\vbox{\kern8pt \hbox{$\displaystyle #1$} \kern8pt}
\kern8pt\vrule}\hrule}}

\begin{document}
 
\thispagestyle{empty}

\begin{flushright} 
\end{flushright}

\vspace{1cm}
\setcounter{footnote}{0}

\begin{center}

{ \Large\bf Effective Quantum Field Theory for the Thermodynamical
Bethe Ansatz}

\vspace{20mm} 

Ivan Kostov
  
 \bigskip
 
{\it Institut de Physique Th\'eorique, DRF-INP 3681, C.E.A.-Saclay, \\
	     F-91191 Gif-sur-Yvette, France} \\[5mm]
\end{center}

\vskip18mm \noindent{ We construct an effective Quantum Field Theory
for the wrapping effects in 1+1 dimensional models of factorised
scattering.  The recently developed graph-theoretical approach to TBA
gives the perturbative desctiption of this QFT. For the sake of
simplicity we limit ourselves to scattering matrices for a single
neutral particle and no bound state poles, such as the sinh-Gordon
one.  On the other hand, in view of applications to AdS/CFT, we do not
assume that the scattering matrix is of difference type.  The
effective QFT involves both bosonic and fermionic fields and possesses
a symmetry which makes it one-loop exact.  The corresponding path
integral localises to a critical point determined by the TBA equation.
}

\newpage
\setcounter{footnote}{0}
%

%

\newpage

\section{Introduction}

In 1+1 dimensional field-theoretical models with factorised
scattering, the momenta and the energies of the particles forming a
multi-particle excitation are not changed by the Hamiltonian
evolution.  This property makes possible to use the notion of a
particle with given rapidity even after the theory is confined on a
circle with asymptotically large circumference $R$.  Asymptotically
large signifies much larger than the correlation length, so that the
exponential effects can be neglected.  The asymptotic energy spectrum
is computed by solving the Bethe-Yang equations which determine the
allowed values of the momenta compatible with the periodic boundary
conditions.

 The leading order exponential corrections to the energy of the ground
 state, known as wrapping corrections, come from virtual particles
 wrapping the space circle.  To compute the energy spectrum at finite
 volume $R$, one should evaluate the sum over all possible wrapping
 processes.  An efficient and elegant way to do that, suggested by
 Alexey Zamolodchikov \cite{Zamolodchikov:1989cf,Zamolodchikov:1991vx,
 Zamolodchikov:1991vg}, is to compactify also the time circle at
 asymptotically large distance $L$ and formulate the problem in the
 mirror channel where the space and the time are interchanged.  Then
 the sum over the virtual wrapping particles, which become real
 particles in the mirror channel, can be performed with the techniques
 of the Thermodynamic Bethe Ansatz (TBA) \cite{YY}.  The sum is
 saturated by a saddle point determined by a set of non-linear
 integral equations for the particle densities known as TBA equations.
 This strategy works also in the AdS/CFT integrability
 \cite{Integrability-overview-2012}, where there is no explicit
 Lorentz invariance.  In this case the dynamics in the physical and in
 the mirror channels look differently and one speaks of physical and
 mirror theories.

Although the TBA equations are derived for the thermodynamical state
involving infinite number of particles, they are able to reproduce the
L\"uscher corrections \cite{Luscher:1985dn} in the limit of
asymptotically large $R$ where the wrapping particles are scarce and
the thermodynamical arguments are not justified.  This suggests that
the TBA equations are in fact exact identities for some statistical
theory describing the grand canonical ensemble of virtual particles in
the mirror channel.  Such a statistical description, which technically
represents an exact cluster expansion, was obtained by several authors
in \cite{2004JPSJ...73.1171K,2002JMP....43.5060K,2001JMP....42.4883K,
2001PhRvE..63c6106K,Kostov:2018ckg,Kostov:2018dmi}.  The latter proved
to be a useful tool to solve some concrete problems \cite{Vu:2018iwv,
vu2019cumulants}.

Another possible strategy to go beyond the thermodynamics is to
consider the particle densities as fluctuating fields and evaluate the
quantum corrections to the TBA equation.  Woynarovich
\cite{Woynarovich:2004gc} found that the gaussian fluctuations lead to
a non-extensive $O(1)$ term in the free energy.  Pozsgay
\cite{Pozsgay:2010tv} then pointed out that in case of periodic
boundary conditions second subleading contribution coming from the
Gaudin measure which cancels the first one.  Very recently Jiang,
Komatsu and Vescovi \cite{Jiang:2019xdz} proposed a path-integral
formulation of the TBA partition function based on an action which,
importantly, differs from the Yang-Yang potential, the most natural
candidate for such an action.\footnote{ The Yang-Yang potential was
introduced in \cite{YY} as a formal tool to prove the existence of the
solutions of Bethe equations by variational principle.  In the last
decade it attracted the attention with its connection, discovered by
Nekrasov and Shatashvili \cite{Nekrasov:2009aa}, to the instanton
counting of four-dimensional gauge theories.  } The path integral
proposed in \cite{Jiang:2019xdz} does generate an $O(1)$ term through
the gaussian fluctuations, but this can be easily cured by introducing
an extra fermionic field which generates the Gaudin measure.

As a matter of fact, it is straightforward to formulate an effective
QFT having as perturbative series the exact cluster expansion for TBA
in the form given in \cite{Kostov:2018ckg,Kostov:2018dmi}.  The idea
for such a construction is contained in the field-theoretical proof of
the matrix-tree theorem given in the appendis B of
\cite{Kostov:2018dmi}.  This QFT involves a pair of bosons as well as
their fermionic partners and the bosonic part of the action matches
the action proposed in \cite{Jiang:2019xdz}.

In this article we adress the inverse problem, namely to construct
from first principles a conceptually satisfactory effective QFT whose
perturbative description is given by the exact cluster expansion.  To
explain the construction in a simple setup, we consider a scattering
theory for a single species of neutral massive particles and no bound
states.  For such a theory there are no Bethe strings and all
excitations have real rapidities.  The only example of such a theory
is given by the sinh-Gordon model, but in view of possible
applications to AdS/CFT, we do not assume relativistic invariance and
the scattering matrix being of difference type.  We only assume that
the scattering phases in the mirror channel are obtained by analytical
continuation of the scattering phases in the physical channel.

We are not going to construct explicitly the wave functions for
periodic boundary conditions and in this respect our approach has no
significant overlap with the field-theoretical derivation of the TBA
equations using the form factor bootstrap discussed in
\cite{Balog-TBA}.  Instead we will postulate that if the scattering
theory is confined in a finite but asymptotically large volume, there
exists, both in the physical and in the mirror theories, an
over-complete set of normalised multi-particle states with
unrestricted rapidities.  Two distinct states are orthogonal if both
sets of rapidities are on-shell, i.e. if they solve the Bethe-Yang
equations.  If the field theory can be constructed as a continuum
limit of a spin chain, which is so in most of the interesting cases
\cite{Polyakov:1983tt, Ogievetsky:1984pv, Faddeev:1985qu,
Destri:1994bv, Destri:1997yz, paulzinn-thesis,Volin:2010aa,
Teschner:2008ab, Integrability-overview-2012}, these states are the
off-shell Bethe states of the Algebraic Bethe Ansatz.

The paper is organised as follows.  In section
\ref{section:elextastorus} we introduce the Hilbert space and the
operator content of the effective QFT. The Hilbert space is generated
by two families of operators which we call {\it wrapping operators}
because they create particles wrapping the main cycles of the torus.
The wrapping operators associated with the same cycle strictly commute
while the operators associated with two intersecting cycles commute by
the scattering matrix.  The modular invariance leads to operator
constraints which are equivalent to the Bethe-Yang equations.  The
wrapping operators are represented as vertex operators acting in the
Fock space of a pair of bosonic fields.  The two bosonic fields are
the operators for the scattering phases of a mirror particle wrapping
the spatial or the temporal cycles.

In section \ref{sect:FockZ} we represent the partition function on a
torus with one of its periods asymptotically large as the expectation
value in the Fock space of the effective QFT. We use the operator
formulation of the Bethe-Yang equations to transform the sum over
on-shell states into a contour integral.  The integration measure
involves the Gaudin determinant, which is taken into account by an
extra pair of fermionic fields.

In section \ref{sec:pathint} we formulate the path-integral
formultaion for the effective QFT is obtained by introducing two extra
pairs of Hubbard-Stratonovich fields which generate respectively the
bosonic and the fermionic correlators.  We show that the path integral
possesses a fermionic symmetry which gets it localised to the critical
point of the action and renders the theory one-loop exact.  For
periodic boundary conditions the one-loop effects compensate
completely.

In section \ref{section:sinh-G} we consider the example of the
sinh-Gordon model, where we show that the critical point of action is
determined by the discrete Liouville equation.  We conclude with
section \ref{section:conclusions}.  Appendix \ref{app:A} contains our
conventions for the scattering phases in the physical and in the
mirror channels.  In appendix \ref{appendix:C} we derive the operator
representation of the Gaudin determinant.

\medskip

 \section{ Degrees of freedom of the effective QFT }
 \label{section:elextastorus}

\subsection{Space and time wrapping operators }
\la{section:spacetimewrapping}

The effective QFT involves a minimal subset of the degrees of freedom
of the original field theory after the latter is compactified on a
torus with asymptotically large space and time circles.  The Fock
space of the effective QFT is generated by two families of operators,
$\bA(u)$ and $\bB(u)$, associated with the two homotopy cycles, $A$
and $B$, of the space-time torus.  Here we adopt the following
convention for the two cycles,
\be \no A\text{-cycle}={\small \text{temporal circle in the physical
kinematics} =\text{spatial circle in the mirror kinematics } },\\
\no B\text{-cycle} = {\small \text{spatial circle in the physical
kinematics} =\text{temporal circle in the mirror kinematics.} } \ee
To avoid confusion let us stress that the Fock space we construct is
not directly related to the Hilbert spaces of the physical or mirror
theories.  In particular, the mirror transformation is an automorphism
of the Fock space.

\begin{figure}[ht!]
         \centering
                \begin{minipage}[t]{0.9\linewidth}
            \centering
	     \includegraphics[scale=0.79]{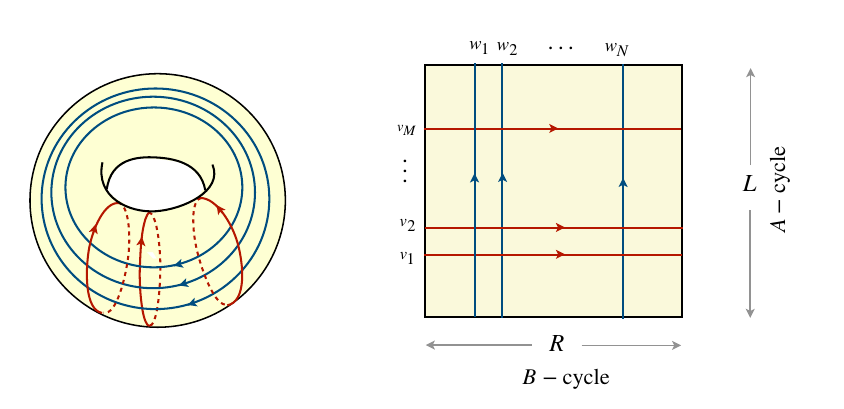}
\caption{ \small Particles wrapping the space and the time circles.
On the left the wrapping particles are depicted as closed paths
winding around the two main circles of the torus.  On the right the
wrapping particles are represented by vertical and horizontal lines on
a periodic rectangle.  The topology requires that each pair of a space
and a time wrapping particles scatter once.  }
\label{fig:multiwrappingspacetime}
         \end{minipage}
 \vskip 0cm
      \end{figure}

The wrapping particles will be represented as non-contractible loops
winding around the two cycles of the torus as shown in fig.
\ref{fig:multiwrappingspacetime}.  This representation is natural in
the microscopic picture based on a spin chain.  When two lines cross,
there is an additional scattering phase factor.  Both types of
wrapping particles can be defined off-shell and their rapidities can
take any complex values.  The $M$-particle off-shell states are
labeled by the non-ordered set of rapidities $\{u_1,..., u_M\}$ and
are assumed to be normalisable but not necessarily orthogonal.  They
are eigenstates of the Hamiltonian with eigenvalues given by the sum
of the energies of the constituent particles.  The on-shell states are
singled out by the periodicity conditions on the rapidities and belong
to a discrete spectrum.

Let us denote by $\bA(u)$ the operator creating a particle of rapidity
$u$ in the {\it physical} kinematics wrapping the $A$-cycle.
Similarly, denote by $\bB(u)$ the operator creating a particle of
rapidity $u$ in the {\it mirror} kinematics wrapping the $B$-cycle.
The operators $\bA$ with real rapidities create real particles in the
physical channel or virtual particles in the mirror channel.
Conversely, the operators $\bB$ with real rapidities create real
particles in the mirror channel or virtual particles in the physical
channel.

An operator defined in the physical theory can be transformed by
analytic continuation $u\to u^{\pm \gamma}$ into an operator defined
in the mirror theory and vice versa.\footnote{Our conventions for the
mirror transformation are summarised in appendix \ref{app:A}.  Our
notations are conform with those used in AdS/CFT integrability.} The
inverse transformation will be denoted by $u^{-\g}$.  For the sake of
clarity we give the list the four types of operators,
  \be\la{fourtypes}
  \begin{aligned}
  \bB(u)\ \ \ \ \qquad &\text{--\ creates
  a mirror   particle wrapping the $B$-cycle,
 }
   \\
  \bA(u^{\g})\ \ \qquad &\text{--\ creates a mirror particle wrapping
  the $A$-cycle} \\
  \bB(u^{-\g}) \qquad &\text{--\ creates  a  physical 
   particle wrapping the $B$-cycle.}
   \\
    \bA(u) \ \ \ \ \qquad &\text{ -- creates a physical particle
    wrapping the $A$-cycle, }
   \end{aligned}
  \ee

The wrapping operators are holomorphic functions of the non-restricted
rapidity variable $u$ and their analytic properties are determined by
those of the S-matrix.  Since the off-shell states are by construction
symmetric under permutations, the operators creating particles
wrapping the same cycle commute.  Morover, the expectation values of
products of wrapping operators of the same kind factorise,
 \begin{align} 
\la{expvalB} \left\langle \prod_{j=1}^M \bB (v_j)
\right\rangle_{\!\!\!  \LR}&= {\langle v_1,..., v_M| \ e^{-
R\tilde\bH}\ |v_1,..., v_M\rangle \over \langle v_1,..., v_M| v_1,...,
v_M\rangle } = \prod_{j=1}^M e^{-R \tilde E(v_j)}, \\
 \la{expvalA}
\left\langle  \prod_{k=1}^N \bA (w_j)\right\rangle_{\!\!\!\LR}& =
{\langle w_1,..., w_M| \ e^{- L \bH}\
|w_1,..., w_M\rangle 
\over
\langle w_1,..., w_M| w_1,..., w_M\rangle } = \prod_{k=1}^N e^{-L
E(w_j)}. 
\end{align}
The expectation values for the virtual wrapping particles are related
to those for the real particles by analytic continuation,
    \be\la{ZFfs}
  \begin{aligned}
  \left\langle \prod_{j=1}^M \bB (u_j^{- \g}) \right\rangle _{\!\!\!
  \LR}= \prod_{j=1}^M e^{ i R p (u_j)} , \quad \left\langle
  \prod_{k=1}^N \bA (u_k^\g) \right\rangle _{\!\!\!  \LR}&=
  \prod_{k=1}^Ne^{ - i L \tilde p(u_k)}\ .
\end{aligned}
 \ee

Now let us consider the expectation value involving both types of
wrapping operators.  For an observer positioned in the physical
channel, the expectation value of $N$ real and and $M$ virtual
particles, illustrated by Fig.  \ref{fig:multiwrappingspacetime}, is
equal to the product of their Boltzmann factors and the scattering
factors between the real and virtual particles,
 \be \left\langle \prod_{j=1}^M \bB (v_j ) \prod_{k=1}^N \bA (w_k
 )\right\rangle_{\!\!\!  \LR} = \prod_{j=1}^ M \prod_{k=1}^N
 S(v^\gamma_j,w_k) \ \prod_{j=1}^ Me^{ - \tilde E (v_j)} \prod_{k=1}^N
 e^{-L E (w_k) }.  \la{vev2FS} \ee

There is an obvious Fock-space representation of the wrapping
operators and the expectation value \re{vev2FS}.  Assume that the two
families of operators satisfy the algebra
\be \la{ABalgebra} \bB(v) \bA(w) = S(v^\g, w)\bA(w)\bB(v), \quad
[\bA(u),\bA(v)]= [\bB(u),\bB(v)]=0 \ee
and introduce the vacuum state $|R\rangle$ and the dual vacuum state
$\langle L|$ so that
  \be \la{ZFalgFock}
    \begin{aligned}
   & \langle L| \bA (u) = e^{ - L E(u)}\, \langle L| , \quad \bB (u)
   |R\rangle = e^{ - R \tilde E(u)}\, |R\rangle.
\end{aligned}
 \ee
With the convention $\< L|R\>=1$, the expectation value \re{vev2FS} is
represented by the scalar product
\be \Big\langle \prod_{j=1}^M \bB (v_j) \prod_{k=1}^N \bA
(w_k)\Big\rangle_{\LR} \defeq \langle L| \prod_{j=1}^M \bB (v_j)
\prod_{k=1}^N \bA (w_k) |R\rangle .  \la{vev2FSFS}
\ee

It is convenient to include in the definition of the Fock-space
expectation value the anti-normal ordering $\ant\ \ \ant$ which
automatically puts all operators $\bB $ on the left of all operators
$\bA $,
\be\begin{aligned} \la{firstoperrepLR} \< \CO\>_{\LR}&\equiv \langle
L| \ant \CO \ant |R \rangle.
\end{aligned}
\ee
With this definition the order of the operators inside the brackets
does not matter.

\subsection{Operator form of the Bethe-Yang equations}

We would like to sum over the on-shell states without actually solving
the Bethe-Yang equations.  This can be achieved by reformulating the
the on-shell conditions on the rapidities as operator constraints.

\begin{figure}[ht!]
         \centering
                \begin{minipage}[t]{0.9\linewidth}
            \centering
             \includegraphics[scale=0.77]{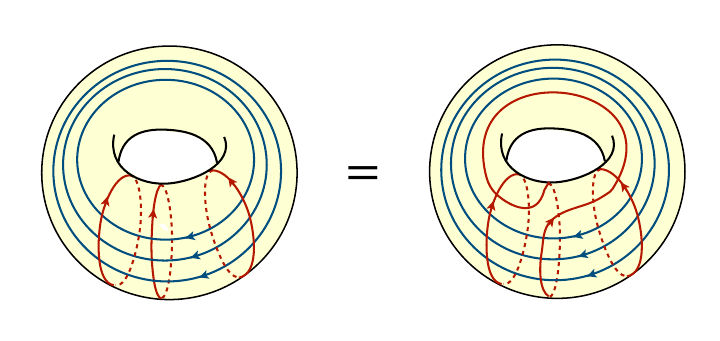}  
	     \caption{ \small Pictorial representation of the
	     quantisation condition \re{BYeqMir} }
\label{fig:Dehn}
         \end{minipage}
 \vskip 0cm
      \end{figure}

The quantisation of the rapidities in the mirror channel is imposed by
the periodicity conditions, which express the invariance of the
expectation value under replacing the cycles $A$ and $ B$ by $A$ and
$B+A$, applied separately for each particle.  In terms of wrapping
operators this signifies that the expectation value is invariant under
replacing $\bB(v_l)$ by $ -\bA(v_l^\gamma)\bB(v_l)$,
\be \la{BYeqMir} \left\langle \Big( 1+ \bA (v_l^\g) \Big) \
\prod_{j=1}^{M} \bB (v_j) \prod_{k=1}^N \bA (w_k)
\right\rangle_{\!\!\!  \LR} =0,\quad \quad l=1, \dots, M. \ee
The minus sign comes from $ \<\bA(v_l^\gamma)\bB(v_l)\>_{\LR} = -1$.
Equation \re{BYeqMir} is equivalent to the Bethe-Yang equations in the
mirror channel,
  \be \la{BYeqs} e^{- i \tilde \phi_l} \equiv e^{- i L \tilde p(v_l)}
  \prod_{j=1}^M \tilde S(v _j, v _l) =-1, \quad l=1, \dots, M, \ee
where $\tilde S$ denotes the S-matrix in the mirror channel,
 \be \tilde S(u,v) \equiv S(u^\g,v^\gamma).  \ee
Intuitively, the insertion of the operator $-\bA(v_l^\gamma )$ has the
effecy of adding an extra piece to the world line of $l$-th mirror
particle: the particle travels around the mirror space before wrapping
the mirror time circle, as shown in fig.  \ref{fig:Dehn}.

In a similar way, the invariance under replacing the cycles $A$ and $
B$ by $A+B$ and $B$ leads to the equations
       \be \la{BYeqPhys} \left\langle \Big( 1+ \bB (w_l^{-\g}) \Big) \
       \prod_{j=1}^M \bB (v_j) \prod_{k= 1}^N \bA
       (w_k)\right\rangle_{\!\!  \LR} =0,\quad \quad l=1, \dots, N,
       \ee
which imposes $N$ constraints for the rapidities of the physical
particles (the Bethe-Yang equatiins in the physical channel)
 \be \la{BYeqsph} e^{ i \phi_l}\equiv e^{ i R p(w_l)} \prod_{k=1}^N
 S(w_l, w_k) =-1, \qquad l=1, \dots, N. \ee

\subsection{Free-field realisation}

The wrapping operators admit a representation as vertex operators for
free fields, similar to for the ZF operators \cite{Lukyanov:1993pn}.
Since we are interested in computing the trace in the mirror channel,
we associate the free fields with the wrapping operators in mirror
kinematics, $\bB(u)$ and $\bA(u^\g)$,
\be\la{Wickbos} \la{Vesp} \bB (u) = e^{- \bvp (u)}, \quad \bA (u^\g) =
e^{ i \bbvp(u)} .  \ee
The algebra \re{ABalgebra} is satisfied if if the two fields have
commutation relations\footnote{Instead of two fields one can speak of
the positive and the negative frequency parts of the same bosonic
field.  This can be done once the concrete form of the scattering
matrix is known.  }\footnote{The branch of the logarithm can be
determined by so that the scattering phase vanishes if one of the
rapidities is sent to indinity.  }
\be
\begin{aligned}
\la{phi2pc} [\bbvp (u), \bvp(v)] =i \log \tilde S( u,v), \qquad [
\bbvp(u), \bbvp(v)] = [ \bvp (u), \bvp (v)] =0.
\end{aligned}
\ee
The action of the two bosonic fields on the Fock vacua is determined
by \re{ZFalgFock}.
  \be \la{ZFalgFockphi}
    \begin{aligned}
  \langle L| \bbvp (u) =\bar \vp^\circ (u) \, \langle L| , \quad \bvp
  (u) |R\rangle = \vp^\circ(u)|R\rangle,
\end{aligned}
 \ee
where we introduced notations for the classical values $ \bar
\vp^\circ $ and $ \vp^\circ $ of the two fields,
  \be \begin{aligned} \la{defcirc}
  \vp^\circ(u)= R \tilde E(u),
\qquad 
 \bar\vp^\circ(u)= L\tilde p(u)  
 .
\end{aligned}\ee
More generally the classical value $\varphi ^\circ$ can be chosen to
be any linear combination of conserved charges densities $\tilde
h_n(u)$ in the mirror theory,
  \be\la{expansioncharges}
  \varphi ^\circ(u) = \sum_n t_n \tilde h_n(u)
  \ee
 with $t_1= R$ and $ \tilde h_1(u) = \tilde E(u)$.  The expectation
 value of operators built from the bosonic fields is defined again by
 \re{firstoperrepLR}, where the anti-normal ordering $\ant\ \ \ant$
 means that all operators $\bvp $ should be put on the left of all
 operators $\bbvp$.

In terms of the gaussian field the on-shell condition for the mirror
$M$-particle states takes the form
\be \la{BYeqMirgaus} \left\langle \Big( 1+ e^{i\bbvp(v_l)}\Big) \
\prod_{j=1}^{M} e^{- \bvp (v_j)} \right\rangle _{\!\!  \LR}=0,\quad
\quad l=1, \dots, M. \ee

   \section{ Partition function on the torus}
    \la{sect:FockZ}

    \subsection{The partition function as a sum over on-shell states}

 The partition function on the torus can be expressed as a thermal
 trace of the evolution operator either in the Hilbert space $\CH$ in
 the physical channel or in the Hilbert space $\tilde \CH$ in the
 mirror channel,
    \be \CZ(L,R)= {\Tr}_{_{ \CH}}\, e^{- L\bH_R}= \Tr_{_{\tilde
    \CH}}\,\, e^{- R \tilde \bH_L} .  \ee
 The first representation is also called $L$-picture while the second
 representation is called $R$-picture.  When $L$ is asymptotically
 large and $R$ is finite, the sum in the $R$-picture is dominated by
 the ground state contribution since the relative contribution of the
 excited states is exponentially small.  The trace in the $R$-picture
 is not accessible for calculation because the spectrum at finite $R$
 is not known.  On the other hand, in the $L$ picture we can replace
 the hamiltonian $\tilde \bH_L$ by $\tilde \bH \equiv \tilde
 \bH_\infty$ whose spectrum is given by the Bethe-Yang equations.
    
The $M$-particle on-shell states in the mirror channel are
characterised by the semi-integer numbers $n_1<n_2<...<n_M$, called
Bethe numbers, which arise in the logarithmic form of the Bethe-Yang
equations
   \be \la{BYeqssl} \tilde \phi_l(v_1,..., v_M) \equiv \< \bbvp(v_l)
   \prod_{j=1}^Me^{-\bvp(v_j)} \> =  2\pi n_l , \quad j=1, \dots, M.
   \ee
The partition function in the $L$-picture is given by the sum over all
Bethe numbers
 \be \la{insertcompleteset} \CZ(L, R) = \sum_{M=0}^\infty \ \sum_
 {n_1<...< n_M} \prod_{j=1}^M e^{-R\tilde E(v_j)} , \ee
with $\vv=\{ v_1,..., v_M\}$   given  implicitly as  the solution of \re{BYeqssl}. 
The Bethe numbers should be all different because of the ``fermionic"
property \re{fermionicproperty} of the S-matrix.

We would like to write this series as a Fock space expectation value.
For that we need first to get rid of the condition that all rapidities
in the sum in \re{insertcompleteset} are different.  This condition
can be relaxed by introducing ``non-physical'' solutions of the
Bethe-Yang equations \re{BYeqssl} such that the rapidities of two or
more particles can coincide \cite{Balog-TBA, 2001PhRvE..63c6106K ,
2001JMP....42.4883K , 2002JMP....43.5060K, Kostov:2018ckg,
Woynarovich:2010wt}.  A stack of $r$ identical Bethe roots $u$ will be
denoted by $u^r$ and $r$ will be called multiplicity.  The generalised
$M$-particle states are characterised by the rapidities and their
multiplicities $\vv^\rr=\{ v_1^{r_1},..., v_m^{r_m}\}$, where $M= r_1
+...+r_m$.  The logarithmic form of the Bethe-Yang equations for such
states is
   \be\begin{aligned} \la{BYeqsslmul}
   \tilde  \phi_l
   & =  2\pi n_l \, , \quad l=1, \dots, m, \\
    \tilde  \phi_l&
   \equiv 
      \< \bbvp(v_l)  \prod_{j=1}^m e^{- r_j\bvp(v_j)}  \>
    =
    \tilde {p}(u_l)L + {1\over i}  \sum_{k(\ne l)}^m r_k\, \log \tilde S
    (u_l,u_k) + \pi (r_l-1) .
    \end{aligned} \ee
Now the sum over ``repulsing'' Bethe  numbers can be expressed as a
non-restricted sum of mode numbers and their multiplicities
\be \la{PrtfnDisc} 
\begin{aligned}
\CZ(L , R)&= \sum_{M=0}^\infty \ \ {1\over M!} \sum_{n_1 , \dots, n_M}
\ \prod_{j<k}^M\(1- \d_{n_j, n_k}\) \prod_{j=1}^M e^{-R\tilde E(u_j)}
\\
&= \sum_{m=0}^\infty {1\over m!} \sum_{r_1,..., r_m\ge 1}\ \sum_{ n_1
,..., n_m } \prod_{j=1}^m {(-1)^{r_j-1}\over r_j}\ e^{-r_jR \tilde
E(u_j)} .
  \end{aligned}
   \ee
The last line is obtained by expanding the product of Kronecker
symbols.   

 \begin{figure}[h!]
	 \centering
	 \begin{minipage}[t]{0.9\linewidth}
            \centering
             \includegraphics[scale=0.45]{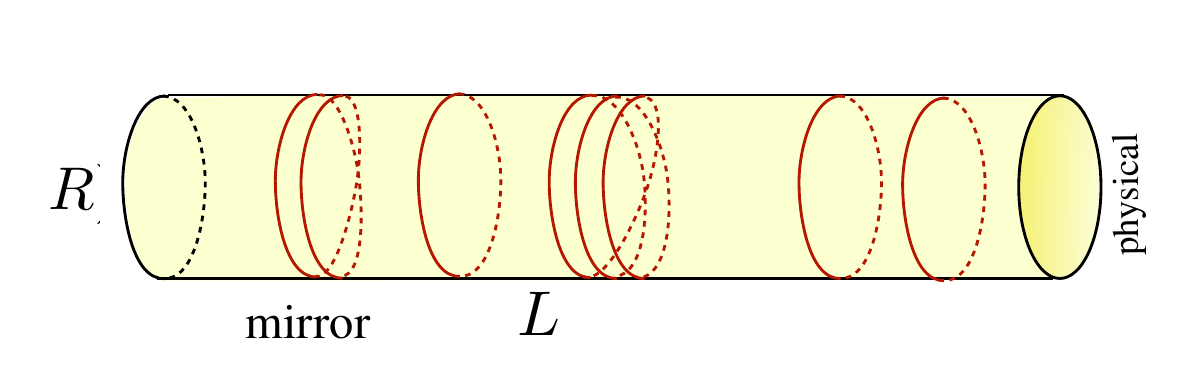} 
	     \caption{ \small Virtual particles wrapping the space
	     multiple times}
\label{fig:multiwrapping}
         \end{minipage}
 \vskip 0cm
      \end{figure}

A Bethe root of multiplicity $r$ can be interpreted as a single mirror
particle wrapping $r$ times the $R$-cycle as shown in fig.
\ref{fig:multiwrapping}.  The operator creating such a particle is
  \be \begin{aligned} \underbrace{
  \includegraphics[height=8ex]{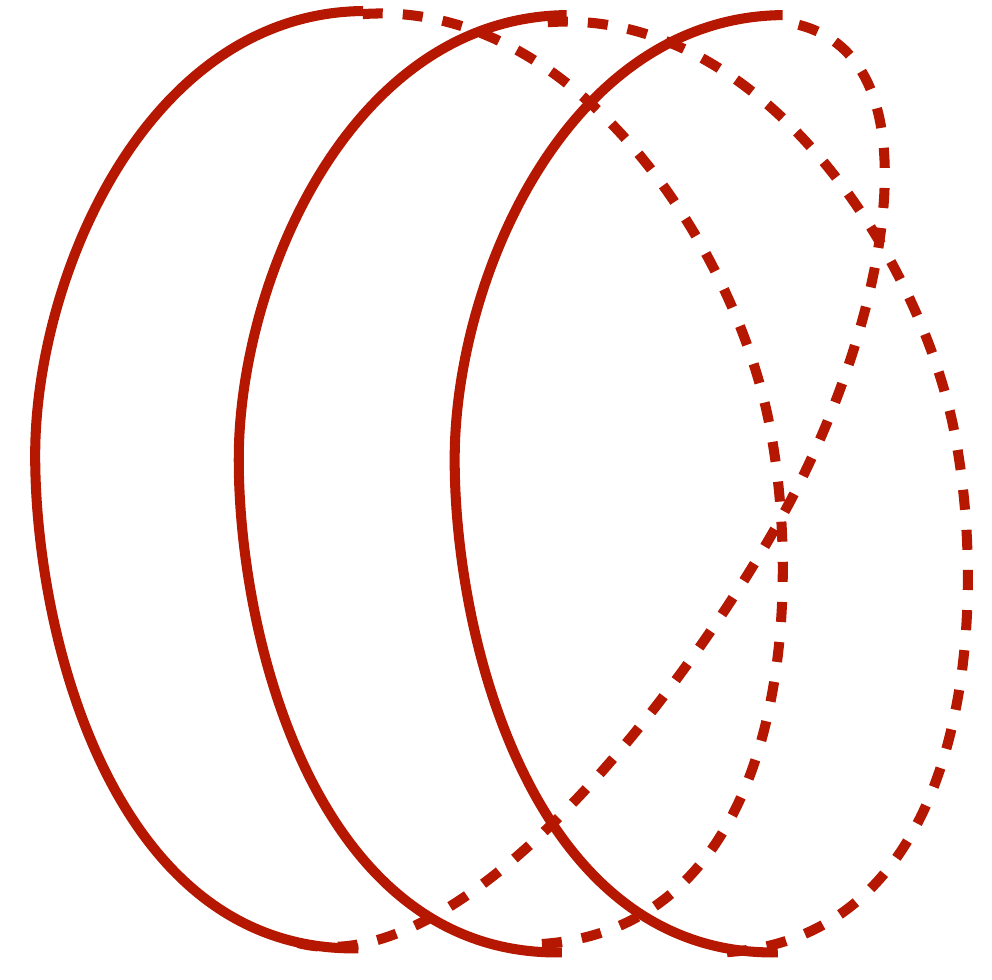}}_{r} \quad =
  {(-1)^{r-1}\over r} \ e^{-r\bvp(u)}.
  \end{aligned}
  \ee
An $r$-wrapping particle scatters with itself $r-1$ times, which
yields a factor $[\tilde S(u,u)]^{r-1} = (-1)^{r-1}$.  The
combinatorial factor $1/r$ counts for the $\IZ_r$ cyclic symmetry of
the $r$-wrapping loop.

\subsection{Rewriting the sum as a contour integral}

Our goal is to perform the sum over the on-shell states without
solving the Bethe-Yang equations.  This can be achieved by
transforming the sum over the Bethe numbers in eq.  \re{PrtfnDisc}
into a multiple contour integral.  Let us first consider the sector
$\tilde \CH_{L, \rr}$ of the Hilbert space $ \tilde\CH $ in the mirror
channel, spanned by all generalised on-shell states of the form
$\uu^\rr=\{u_1^{r_1},..., u_m^{r_m}\}$ with fixed wrapping numbers
$\rr=\{r_1,..., r_m\}$.  The on-shell states belonging to $\tilde
\CH_{L, \rr}$ are characterised by $m$ unrestricted Bethe numbers
$n_1,..., n_m$.

The thermal trace restricted to this sector can be expressed as an
$m$-fold contour integral involving the $m$ scattering phases defined
by \re{BYeqsslmul},
\be \begin{aligned} \Tr_{_{\tilde \CH_{L, \rr}}}e^{- R \tilde \bH_L}&=
\la{multicontour} \sum_{n_1,...,n_m} \prod_{j=1}^m \ e^{-r_jR \tilde
E(u_j)} \\
&= \oint\limits_{\IR ^m} \ \prod_{j=1}^m { d\log [1+ e^{i\tilde
\phi_j} ] \over 2\pi i} \ e^{- r_j R \tilde E(u_j)}, \end{aligned}\ee
with each contour of integration encircling the real axis
anticlockwise.  Expanding the differentials we write, more explicitly,
\be \begin{aligned} \la{intermedint} \text{rhs of }
\re{multicontour}=(-i)^N \oint\limits_{\IR ^m} \prod_{j=1}^m { du_j
\over 2\pi i} { e^{- r_j R \tilde E(u_j)}\ \over 1+ e^{-i\tilde
\phi_j}} \, \tilde G(\uu^{\rr} ) , \end{aligned}\ee
where the last factor is the Jacobian of the change of variables from
the phases to the rapidities known also as Gaudin determinant,
\be
\la{defgaudin}
 \begin{aligned}
\tilde G(\uu^\rr) &=\det\tilde G_{kj},
\quad
\tilde G_{kj}&= {\p\tilde \phi_j(\uu^\rr)\over \p u_k}.
 \end{aligned}\ee

Now we are going to represent the integrand in \re{intermedint} as a
Fock-space expectation value.  For that we have to give an operator
representation of the Gaudin determinant.  As we show in appendix
\ref{appendix:C}, this can be achieved by introducing extra fermionic
fields $ \bbpsi(u) $ and $ \bpsi(u)$ defined by
\be
\begin{aligned}
\la{psi2pc} [\bpsi (u), \bbpsi(v)]_+ &=-i \log \tilde S( u,v), \qquad
[ \bbpsi(u), \bbpsi(v)] _+ = [ \bpsi (u), \bpsi (v)] _+=0.  \\
  \langle L| \bbpsi (u) &= 0 \,  , \quad
  \bpsi (u) |R\rangle =  0.
\end{aligned}
 \ee
The expectation value $\langle \ \ \rangle_\LR$ is extended to
operators involving fermionic fields again by the anti-normal product
convention.  The fermionic formula for the Gaudin determinant, which
will be the basic technical tool for constructing the effective QFT,
reads
    \begin{align}
 \tilde G(\uu^\rr) \, \prod_{j=1}^m e^{- R\, r_j \tilde E(u_j)}=
 \left\langle \prod_{j=1}^m \[ \bbvp'(u_j) - r_j \, \bbpsi(u_j)
 \bpsi'(u_j) \] e^{-r_j \bvp (u_j) }\right\rangle_{\!\!\LR} .
 \la{GaudinasFF}
 \end{align}
 With the help of the fermion ic formula we write the rhs of
 \re{multicontour} as
\be
\la{operepsum}
 \begin{aligned}
 \text{rhs of } \re{multicontour}= \left\langle\ \oint\limits_{\IR^m }
 \prod_{j=1}^m { du_j \over 2\pi i} \ { i \bbvp'(u_j) -i r_j \,
 \bbpsi(u_j) \bpsi'(u_j) \over 1+e^{i\bbvp(u_j)}} e^{-r_j \bvp (u_j)
 }\right\rangle .  \end{aligned}\ee
Performing the sum over the multipliticites in \re{GaudinasFF} with
the appropriate weights, we obtain the desired Fock-space
representation of the partition function,
 \be\begin{aligned} \la{firstoperrep} \CZ(L,R) & =\left\langle \hat
 \bO \right\rangle_{\!  \LR},
 \end{aligned}
\ee
with the operator $\hat \bO$ defined by
\be \begin{aligned} \la{defhatomega} \hat\bO &\equiv \exp\oint_{\IR}
{du\over 2\pi i} \(  i \bbvp'(u) \log(1+ e^{ -\bvp(u)}) + {i\, \bbpsi
(u) \bpsi'(u)\over 1+e^{\bvp(u)} } \) {1\over 1+ e^ {-i \bbvp (u)} } \\
&= \exp\oint_{\IR} {du\over 2\pi i}
\[\log(1+
 e^{-\bvp (u)} ) 
 \p_u \log (1+e^{ i\bbvp(u) })
  + { i\, \bbpsi(u) \bpsi'(u) \over (1+ e^{ \bvp (u)})(1+ e^{-i\bbvp(u)})} 
 \]  ,
 \end{aligned}
\ee
the contour of integration encircling the real axis anticlockwise.

Remarkably, the integrand is symmetric with respect of exchanging
$i\bbvp$ and $\bvp$.  The asymmetry is introduced by the contour of
integration which encircles the zeroes of $1+ e^{i\bbvp}$ only and not
those of $1+ e^{ \bvp}$.

 \subsection{   Continuous spectrum approximation}
 \label{section:FFR2}

The operator representation \re{firstoperrep} is justified only when
the bare expectation value $\bar\vp^\circ =\<\bbvp\> $ is
asymptotically large, $\bar\vp^\circ = L \tilde p(u)$.  Then one can
replace in \re{defhatomega} with exponential accuracy
 \be
 \log(1 + e^{   i  \bbvp (u)})\ \to\ 
 \begin{cases}
      i \bbvp(u) & \text{ if } \ \Im u<0, \\
   \quad   0 & \text{if} \ \Im u>0\, 
\end{cases}
  \ee
and
 \be
 {1\over 1 + e^{  -i  \bbvp (u)}}\ \to\ 
 \begin{cases}
   0   & \text{ if } \ \Im u>0, \\
   1 &\  \text{if} \  \ \Im u<0\, .
\end{cases}
  \ee
Hence the contour integral in \re{defhatomega} can be approximated by
a linear integral slightly above the real axis, with an extra minus
sign because of the orientation of the contour, and the operator
representation \re{firstoperrep} simplifies to
 \be\begin{aligned} \la{secondoperrepl} \encadremath{ \CZ(L,R)
 =\left\langle \check \bO \right\rangle_{\LR}, \quad \check \bO \equiv
 \exp\int\limits _{-\infty}^\infty {du\over 2\pi }
\[\log(1+e^{-\bvp(u)} )
  \bbvp' (u ) - { \bbpsi(u) \bpsi'(u)\over 1+e^{\bvp(u)} } \].
  }\end{aligned}
\ee

The expectation value in \re{secondoperrepl} generates the sum over
all solutions of the Bethe-Yang equations.  Since the scattering
phases are of order of $L$, the rapidities are spaced at $\sim 1/L$
and the sum can be approximated by an integral up to exponentially
small in $L$ corrections.
  
 The graphs in the combinatorial approach of \cite{Kostov:2018ckg}
 appear as Feynman diagrams when the expectation value in
 \re{secondoperrepl} is evaluated perturbatively starting with the
 classical values $\bar\vp^\circ$ and $\vp^\circ$.  Denoting by
 $\bvp_q$ and $\bbvp_q$ the quantum part of the bosonic fields,
  \be \bvp(u) = \vp^\circ(u) +\bvp_q(u), \quad \bbvp(u) = \bar
  \vp^\circ(u) +\bbvp_q(u), \ee
 the Feynman graphs are composed of vertices generated by the
 expansion of the non-polynomial potential $\log(1+ e^{-\vp^\circ
 -\bvp_q })$, and propagators given by
   %
   \be \la{propagators} \< \bbvp'_q(u) \bvp_q (v)\>_{\LR} =-  \tilde
   K(u,v), \quad \left\langle   \bbpsi '(u) \bpsi(v)
   \right\rangle_{\LR}=- \tilde K(u,v) , \ee
   where $\tilde K$ is the   scattering kernel in the mirror theory
  \be \la{deftildeK} \tilde K (u,v) =
  {1\over i} \partial_u \log \tilde S(u, v)
.
\ee
 The propagators are in general not symmetric and should be represented by
 oriented lines.

  \begin{figure}[h!]
          \centering
         \begin{minipage}[t]{0.75\linewidth}
 \includegraphics[scale=0.25]{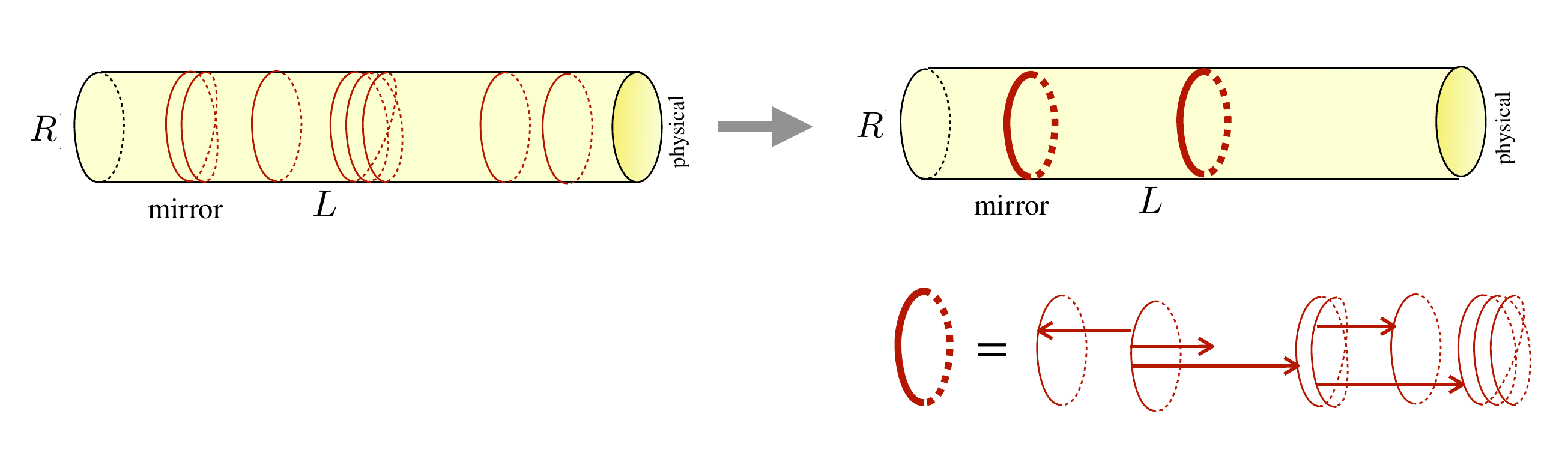} 
  \caption{\small Wrapping particles put in a large box of length $L$
  form a non-ideal gas with interaction $\sim 1/L$.  The free energy
  is a sum over non-interacting clusters of wrapping particles.  The
  perturbative series of the expectation value \re{secondoperrepl}
  gives the exact cluster expansion to all orders.}
   \label{fig:clusters}
    \end{minipage}
         \end{figure}

As in any perturbative QFT, the free energy $\CF(L,R)=\log \CZ(L,R)$
is equal to the sum of all connected Feynman graphs, which are either
trees or trees with one cycle.  For periodic boundary conditions the
fermionic and the bosonic cycles compensate each other and only the
trees contribute to the free energy.  These trees describe clusters of
wrapping particles sketched in Fig.  \ref{fig:clusters}.  The
generating function for the trees satisfies a non-linear integral
equation, which is nothing but the TBA equation for a theory with
diagonal scattering \cite{Kostov:2018ckg}.

\subsection{Excited states in the physical channel}
\la{section:excitedstate}

Let us consider the more interesing case of an excited state with
rapidities $\ww=\{w_1,...,.  w_M\}$ in the physical channel.  The
excited states in finite volume have the same particle structure as in
the asymptotically large volume, but the rapidities of the particles
satisfy the so called ``exact Bethe equations", which take into
account the wrapping effects.
  
The excited state partition function is obtained by performing in the
expectation value \re{vev2FSFS} the sum over all on-shell rapidities
$\vv$.  Equivalently, one can insert in the expectation value in
\re{secondoperrepl} $N$ physical wrapping operators with rapidities
$\ww=\{w_1,...,.  w_M\}$,
     \be
   \begin{aligned}
    \la{operatorbisexphi} & \mathcal{Z}(L ,R, \ww) = \bigg\langle
    \prod_{j=1}^N \bA(w_j) \ \check \bO \bigg\rangle _{\!\!  \LR}\, ,
    \quad \bA(w^\g ) \equiv \exp [-i \bbvp(w)].
    \end{aligned}
  \ee
The rapidities $\ww$ must solve the exact Bethe equations in the
physical channel, which is equivalent to imposing the operator
constraint
 \be \bigg\langle \( 1+ \bB(w_l ^{-\g})\) \prod_{j=1}^N \bA(w_j) \
 \check \bO\ \bigg\rangle _{\!\!\LR}=0, \quad l=1,..., N,
 \la{exactBAE} \ee
 with $\bB(w)\equiv  e^{- \bvp(w)}$.

The excited state contributes to the free energy directly by the
energies of the $N$ particles as well as indirectly by modifying the
Bethe-Yang equations for the rapidities of the virtual particles in
the mirror channel.  This is taken into account automatically by the
expectation value.  On the other hand, the mirror wrapping particles
from the thermal ensemble backreact by deforming the quantisation
condition for the rapidities $\ww$.  Therefore if we first compute the
partition function \re{operatorbisexphi} and then determine the exact
Bethe roots $\ww$ from \re{exactBAE}, the result may be incorrect.
 
  The proper way to introduce the excited state is as a contour
  integral which automatically imposes the on-shell condition
  \re{exactBAE}.  That is, the product of operators inserted in the
  expectation value \re{operatorbisexphi} should be replaced by a
  multiple contour integral around the exact Bethe roots.  The
  integrand of the $N$-fold contour integral contains a Jacobian which
  can be expressed in terms of the free fermions as in \re{operepsum},
  but with $\bvp$ and $i \bbvp$ interchanged.  This results in
  replacing in the expectation value \re{operatorbisexphi}
\be \begin{aligned} \la{multicontourexit} \prod_{j=1}^N e^{  i \bbvp(w
_j)} \to \ &\ {(-1)^N\over N!} \oint\limits _{\ \ \ww^{-\g}}
\prod_{j=1}^N {du_j\over 2\pi i} \, e^{  i \bbvp(u _j)} \ { \bvp'(u_j)
- \bbpsi(u_j)\bpsi'(u_j) \over 1+ e^{ \bvp(u_j)}} .  \end{aligned}
\ee
However there is another, more elegant way to impose the on-shell
condition, namely by lifting the on-shell condition in the exponent,
which allows to write it as a single contour integral,
\be
 \begin{aligned}
   \la{multicontourop} \prod_{j=1}^N e^{  i \bbvp(w _j)} \to \ &\ \exp
   \Big[ \oint\limits _{\ww^{-\g}} {d u\over 2\pi i} \ ( i \bbvp(u))\
   { \bvp'(u)- \bbpsi(u) \, \bpsi'(u)\over 1+ e^{ \bvp(u)}} \Big] .
  \end{aligned} 
  \ee
The equivalence of \re{multicontourexit} and \re{multicontourop} is
intuitively clear, but we do not know a simple formal proof.  It can
be established order by order by expansing the exponential and and
evaluating the residues for each term.

 After inserting the rhs of \re{multicontourop} in
 \re{operatorbisexphi} and integrating by parts we see that, in accord
 with the original Dorey-Tateo prescription, the expression for the
 excited partition function is obtained by extending the integration,
 originally along the real axis, by a contour encircling the roots
 $\ww^{-\g}$,
\be
\begin{aligned}
 \la{excitedZexplicit} \mathcal{Z}(L ,R, \ww) &= \bigg\langle \exp
 \int \limits_{\IR\cup \ww^{-\g}} {du\over 2\pi } \(\p_u \bbvp \
 \log\(1+e^{-\bvp} \) - { \bbpsi \, \bpsi'\over 1+e^{\bvp} } \)
 \bigg\rangle _{\!\LR}.
  \end{aligned} 
      \ee

It happens that in the case of periodic boundary conditions the
na\"ive prescription \re{operatorbisexphi}-\re{exactBAE} gives the
correct result.  This is an accident.  In more general situations, as
in the case of open boundaries in the mirror channel, the operators on
the r.h.s. of \re{multicontourexit} or \re{multicontourop} contribute
non-trivially to the subleading order, as it was recently discovered
in \cite{Jiang:2019xdz,Jiang:2019zig}.  In the periodic case the
subleading contributions cancel completely.

\section{ Path integral  and localisation}
 \la{sec:pathint}

The most transparent formulation of the effective QFT is given by a
path integral.  As we will see later, the integration measure does not
need any regularisation.

 To formulate the path integral for the effective QFT one should add,
 in addition to the fields $\vp, \bar\vp, \psi, \bar\psi$, two extra
 pairs of fields, ${\color{red}\rho, \bar\rho} $ and $
 {\color{blue}\vartheta , \bar\vartheta }$, which define respectively
 the bosonic and fermionic propagators.  The expectation value
 \re{secondoperrepl} is then given by
\be
 \begin{aligned}
    \la{pathintPF} Z(L,R)&= \int\mathcal{D} [\text{fields}] \
    e^{-\mathcal{A}[\text{fields}]}, \\
-\mathcal{A}[\text{fields}]&= \int\limits _{\!\!  -\infty}^\infty
{du\over 2\pi } \left( \log(1+ e^{ -\varphi })\, \p_u\bar\varphi -
{\bar \psi\, \p_u \psi \, \over 1 +e^{ \varphi }} + ( \bar\varphi
-\bar\varphi ^\circ ) {\color{red} \rho }+ {\color{red} \bar\rho }
(\varphi - \varphi ^\circ) + {\color{blue} {\color{blue}} \bar
\vartheta } \psi+ \bar\psi {\color{blue}\vartheta } \right) \ \ \ \ \\
  & - \int\!\!\!\!  \int \limits _{\!\!\!\!\!  -\infty}^\infty
  {du\over 2\pi }{dv\over 2\pi } \ \tilde \phi (u, v)\({\color{red}
  \rho} (u) {\color{red}\bar\rho} (v) + {\color{blue} \bar \vartheta }
  (v) {\color{blue} \vartheta } (u) \right) ,
\end{aligned}
\ee
where $\tilde \phi(u,v)$ denotes the scattering phase in the mirror
channel,
 \be \la{defscatphase} \tilde \phi(u,v) = {1\over i} \log \tilde
 S(u,v) .  \ee

As the action of the path integral \re{pathintPF} is linear in the
fields $\bar \vp$ and $\bar \rho$, we write it in the form
 \be \CA = -\int {du\over 2\pi } \ \bar\varphi ^\circ {\color{red}
 \rho } +\int {du\over 2\pi } \( \bar \vp\, \CA _{\bar\vp} + {\color{red}
 \bar \rho } \, \CA_{\color{red} \bar \rho } \)+ \text{fermions}.
  \ee
The fields $\bar \vp$ and $\bar \rho$ play the role of Lagrange
multipliers and can be integrated out, giving functional
delta-functions.  The path integral thus localises to the critical
point $\CA _{\color{red} \bar \rho} = \CA_{\bar\vp } =0$,
\be\begin{aligned} \la{eqsvprho} \CA _{\color{red} \bar \rho } =0\
\Rightarrow \qquad \vp(u) &= \vp^\circ(u) + \int {dv\over 2\pi }
\redr(v)\, \tilde\phi (v,u) , \\
 \CA_{\bar\vp }=0 \ \Rightarrow 
 \qquad 
  \redr(u)&= \partial_u \log(1+ e^{-\vp(u)}).
\end{aligned}
\ee
After excluding ${\color{red}\rho}$ and integrating by parts, one
obtains the TBA equation for the pseudo-energy $\varepsilon(u) \equiv
\vp(u)$,
  \be \la{TBA0} \vp(u) = \vp^\circ(u) - \int{dv\over 2\pi} \log(1+
  e^{- \vp(v)})\ \tilde K(v, u).  \ee
Furthermore the gaussian fluctuations of the bosons and the fermions
completely cancel and the partition function is given by the exponent
of the critical action
 \be\la{TBA1} \CF(L, R) = - \mathcal{A} _{\text{crit}} = L
 \int_{-\infty}^\infty {du\over 2\pi} \ \tilde p'(u) \ \log(1+ e^{-
 \vp(u)}) .  \ee
This paraphrases Pozsgay's argument \cite{Pozsgay:2010tv} about the
absence of an $O(1)$ term in the free energy.

The localisation implies a stronger statement, namely that there are
no $1/L$ corrections to the free energy, see e.g.
\cite{Klassen:1989ui}.  The localisation of the path integral is
generically a consequence of a fermionic symmetry of the measure.  In
our case the fermionic symmetry is generated by the nilpotent
operators
 \begin{align}
    Q(u) & = \bar\psi {\delta\over\delta \varphi }+\bar\varphi
    {\delta\over \delta\psi } + {\color{red} \bar\rho} {\delta\over
    \delta {\color{blue} \vartheta} } + {\color{blue}\bar \vartheta}
    {\delta\over \delta {\color{red}\rho} } .  \\
\bar Q(u) &= \psi {\delta\over\delta \bar \varphi }+ \varphi
{\delta\over \delta \bar\psi } + \rho {\delta\over \delta
{\color{blue}\bar \vartheta} } + \theta {\delta\over
\delta{\color{red} \bar \rho } }
\end{align}
satisfying the algebra $Q^2 =\bar Q^2=0$, \ $Q\bar Q=\bar Q Q=2$.  The
action in \re{pathintPF} takes the form of a sum of a term which does
not produce quantum effects and a $Q$-exact ``localisation term'',
 \begin{align}
\la{susyaction} \mathcal{A} &= \mathcal{A}^\circ + \int {du\over 2\pi}
\, \mathrm{Q}(u) \mathcal{B} , \qquad \mathcal{A}^\circ =-\int
{du\over 2\pi} \, \bar \varphi ^\circ {\color{red}\rho} , \quad
\mathcal{B} = \bar Q\CA.    
   \end{align}
The odd functional $\CB$ 
 is given by the integral
 \begin{align}
\mathcal{B} = -   \int {dv\over 2\pi} \Big(   \psi  ' \log(1+e^{-\varphi
}) +\psi {\color{red}\rho} + {\color{blue}\vartheta}(\varphi -\varphi
^\circ) \Big)  + \int { du\over 2\pi}{dv \over 2\pi}  \
{\color{red}\rho} (v)    \tilde\phi(v,u )   {\color{blue} \vartheta}(u).
   \end{align}

The path integral with such an action localises at its critical point.
The standard localisation argument\footnote{See e.g. the recent review
\cite{Pestun:2016qko}.} is that the second term can be multiplied by
any constant $t$ without changing the path integral:
\be \la{localisAction-1} \mathcal{A} \to \mathcal{A}^\circ + t
\mathrm{Q}\mathcal{B} .  \ee
 The modified path integral
   \be \mathcal{Z} _t = \int e^{- \mathcal{A}^\circ - t
   \mathrm{Q}\mathcal{B}} \ee
   does not depend on the coupling $t$,
  \be {\partial \mathcal{Z} _t\over \partial t} = \int e^{ -
  \mathcal{A}^\circ - t \mathrm{Q}\mathcal{B} } (- \mathrm{Q}
  \mathcal{B} )= \int \mathrm{Q} \left( e^{ - \mathcal{A}^\circ - t
  \mathrm{Q}\mathcal{B} } \mathcal{B} \right) =0, \ee
where it is used the fact that the measure is $Q$-invariant to perform
the integration by parts.  Taking the limit $t\to\infty$ one obtains
\be \mathcal{Z} = e^{ - \mathcal{A}^\circ
}\Big|_{\mathrm{Q}\mathcal{B} =0}.  \ee
The condition $\mathrm{Q}\mathcal{B} =0$ selects the critical point
\re{eqsvprho}.
  
The above argument holds also in the case of an excited state in the
physical channel, in which case the  action contains an extra term
$i \bar\vp(w_1^{-\g})+ ...  + i\bar\vp(w_N^{-\g})$ and critical point
determined by the TBA equation in presence of an excited state,
  \be \la{criticalexcited} \vp(u) = \vp^\circ(u) -i \sum _{k=1}^N
  \tilde\phi ( u , w^{-\g}_k) - \int_{-\infty}^\infty {dv\over 2\pi}
  \log(1+ e^{- \vp(v)}) \tilde K(v,u)\, .  \ee
Here we used that, thanks to the localisation property, the functional
variables can be considered as holomorphic functions of the variable
$u$ and thus can be analytically continued in a strip around the real
axis.  The critical action is given by the well known expression for
the excited state free energy
  \begin{align} \la{TBA0ex} \CF(L,R, \ww) &= -L \sum _{j=1}^N E(w_j) +
  \int{ du\over 2\pi}
  \log\(1+e^{-\vp(u)}\) \p_u\bar\vp^\circ(u).
  \end{align}

In the case of open boundary conditions one obtains a one-loop exact
path integral, with the gaussian fluctuations contributing to the
subleading term in the free energy.  The one-loop contribution then is
given by the sum over the trees with a cycle in the graph expansion
developed in ref.  \cite{Kostov:2019fvw}.

\section{Example: the  Sinh-Gordon model }
\la{section:sinh-G} \def\YY{{Y\!\!\!  \!Y}}

For concrete integrable models of QFT, the path integral
representation \re{pathintPF} can be transformed further to reveal
more of the integrable structure.  Here we examine the Sinh-Gordon
model, described by the Euclidean action \be \mathcal{A} = \int d^2 x
\left[ {1\over 4\pi} (\nabla\phi)^2 + {2\mu \over \sin \pi b^2}
\cosh ( 2 b\phi)\right].  \ee
The model depends on the dimensonless parameter $b$ and the coupling
constant $\mu$ of dimension $[\mathrm{mass}]^{1+ b^2}$.  This is the
simplest non-compact integrable model.  The thermodynamic Bethe Ansatz
for the sinh-Gordon model was formulated by Al.  Zamolodchikov
\cite{Zamolodchikov:2000kt} and S. Lukyanov \cite{Lukyanov:2000jp},
who derived functional equations for the TBA pseudo-energy and related
functions.  More rigorous approach based on an integrable lattice
regularization was developped in \cite{Bytsko:2006aa,Teschner:2008ab}.
Some recent developments concern the computation of the one-point
functions \cite{Negro:2013wga} and diagonal form factors
\cite{Bajnok:2019yik}.
 
As a scattering theory, the sinh-Gordon model involves a single
massive relativistic particle whose mass $m$ is determined by the
coupling constant $\mu$.  If the energy and the momentum are
parametrised by the rapidity $u$ as
\be
\la{ShGDEp}
p(u)= m \sinh \pi u, \ \ E(u)= m \cosh\pi u,
\ee
then the two-particles S-matrix is given by a product of two
``universal'' S-matrices,
\be 
\la{SshSuni}
S(u ) = S_0(u+i a/2)S_0(u- i a/2) , \qquad S_0(u) = \frac{e^{\pi
u}-i}{e^{\pi u}+i} , \ee
with the shift parameter $a$ related to the dimensionless parameter
$b$ by
\be
 a\equiv (1-b^2)/(1+b^2 )
 .
 \ee
 The ``universal'' scattering  kernel  
  \be
  \la{K0def}
K_0(u, v ) \equiv  -i \p_u \log S_0(u)= {\pi  \over \cosh \pi (u -v)}
 \ee
  satisfies the difference equation
 \be
 \la{KoDiLa1}
  K_0 (u +i/2)  +   K_0 (u -i/2)   = 2\pi  \d(u). 
  \ee

  It proves to be very helpful to represent the shifts of the
  rapidities as the result of the action of powers of the analytic
  difference operator
   \be \mathbb{D} = \exp\left({i\over 2} {\p\over \partial u}\right).
   \ee
Then the relations  \re{SshSuni} and  \re{KoDiLa1}
can  be written as
\be \la{operatorrelations} S(u) = e^{ ({\ID^{a} + \ID^{-a}}) \log
S_0(u)} = S_0(u) ^{\ID^{a} + \ID^{-a}}, \qquad (\mathbb{D}
+\mathbb{D}^{-1})K_0 (u) = 2\pi \d(u).  \ee
 Formally the sinh-Gordon scattering kernel $K= - i \p_u \log S =
 (\ID^a+\ID^{-a}) K_0$ represents the operator
\be
\la{Koper}
\IK =  {\mathbb{D}^{a}+\mathbb{D}^{-a}\over \mathbb{D} +\mathbb{D}^{-1}}
 \ee
 acting on the meromorphic functions defined in the rapidity complex
 plane.  We can also invert this operator paying attention to the fact
 that both $\IK$ and $\IK^{-1}$ have non-trivial null spaces.  For
 example, the energy and the momentum \re{ShGDEp} belong to the null
 space of the operator $\ID+\ID^{-1}$.

 Now let us formulate the path integral of the effective QFT for the
 sinh-Gordon model.  Replacing in eq.  \re{pathintPF} the expression
 for the sinh-Gordon scattering matrix in the form
 \re{operatorrelations}, we have
 \begin{align}
 \la{pathintPFloc} \CZ^{\mathrm{ShG}}(L,R)&= \int [\text{fields}] \
 e^{-\mathcal{A}[\text{fields}]}, \\
\mathcal{A}[\text{fields}]&= \int\limits _ {-\infty}^\infty {du\over
2\pi } \left( - \log(1+ e^{ -\varphi })\, \bar\varphi ' - { \bar \psi
\,\psi ' \, \over 1 +e^{ \varphi }} + ( \bar\varphi -\bar\varphi
^\circ ) {\color{red} \rho }+ (\varphi - \varphi ^\circ) {\color{red}
\bar\rho }+ {\color{blue} {\color{blue}} \bar \vartheta } \psi+
\bar\psi {\color{blue}\vartheta } \right) \no \\
  &  - i\int  {du\over 2\pi }{dv\over 2\pi }   
  \(
    {\color{red}\bar\rho} (u)   {\color{red} \rho}(v)-
  {\color{blue}  \vartheta } (v) {\color{blue}  \bar\vartheta  }(u)\right) 
 (\mathbb{D}^ {a} + \mathbb{D}^{-{a}}) \log   S_0(u-v)
\la{ActionA}
\end{align}
with 
 \be \vp^\circ(u)=R m \cosh \pi u, \quad \bar\vp^\circ(u)= L m \sinh
 \pi u.  \ee

 The components of $\rho$ and $\bar \rho$ in the null space of the
 operator $\ID+\ID^{-1}$ impose the asymptotic conditions at infinity
\be
\la{bcvpbvp}
\vp(u)\simeq \vp^\circ(u) ,\quad
\bar \vp(u)\simeq\bar  \vp^\circ(u)\qquad (u\to\infty).
\ee
We will impose instead these boundary conditions by hand and
simultaneously restrict the functional integration in $\rho$ and $\bar
\rho$ to the space orthogonal to the null space of the operator
$\ID+\ID^{-1}$.  We assume that the fluctuating field variables can be
analytically continued in the strip $ |\Im u|<1/2$, so that they can
be acted on by the shift operators $\ID^{\pm 1}$.

The property \re{operatorrelations} of the universal kernel
can be used to simplify the integral.  After a change of variables
  \be \la{changebartilde} {\color{red} \bar \rho}= ( \mathbb{D}
  +\mathbb{D}^{-1}) \chi, \quad {\color{blue}\bar \vartheta}= (
  \mathbb{D} +\mathbb{D}^{-1}) \xi, \ee
 the  action   takes   a quasi-local  form
 \begin{align}
  \la{pathintPFloc1} \mathcal{A}[\text{fields}]& = \int {du\over 2\pi }
  \left( - \log(1+ e^{ -\varphi })\, \bar\varphi ' - { \bar \psi
  \,\psi ' \, \over 1 +e^{ \varphi }} + ( \varphi - \varphi ^\circ ) (
  \mathbb{D} +\mathbb{D}^{-1}) \chi\right) \\
  &+\int {du\over 2\pi } \left( (\bar \varphi - \bar \varphi ^\circ)
  {\color{red} \rho }+ \bar \psi {\color{blue} \vartheta } + \psi (
  \mathbb{D} +\mathbb{D}^{-1}) \xi \right) \no \\
  & -\int {du\over 2\pi }{dv\over 2\pi } \( {\color{red}\rho }
  (\mathbb{D}^ {a} + \mathbb{D}^{-{a}}) \chi + {\color{blue} \vartheta
  } (\mathbb{D}^ {a} + \mathbb{D}^{-{a}}) \xi\right) .
\end{align}
The fields $\rho$ and $\vartheta$ can be integrated out giving the
constraints $\bar\vp =\bar\vp^\circ + (\mathbb{D}^ {a} +
\mathbb{D}^{-{a}}) \chi$ and $\bar\psi= (\mathbb{D}^ {a} +
\mathbb{D}^{-{a}}) \xi$.  As a result the action for the path integral
simplifies to
 \be\begin{aligned}
 \la{actionA}
  \mathcal{A}[\varphi , \chi, \psi,\xi ]= & -\int {du\over 2\pi}
  \p_u\bar\vp^\circ \log(1+ e^{-\varphi }) \\
 & + \int {du\over 2\pi} \left[ \varphi
 (\mathbb{D}^{}+\mathbb{D}^{-1}) \chi- \log(1+ e^{-\varphi
 })(\mathbb{D}^a +\mathbb{D}^{-a})  \chi \right] \\
  &+ \int {du\over 2\pi} \left[\psi (\mathbb{D}^{}+\mathbb{D}^{-1})
   \xi-{ 1 \over 1+ e^{\varphi }} \psi (\mathbb{D}^a
  +\mathbb{D}^{-a})   \xi\right]\,
\end{aligned}\ee
with asymptotic condition $\vp\simeq \vp^\circ$ at $u\to\infty$.  The
action is linear in $\chi$ and the path integral localises to the
critical point which is the solution of the discrete Liouville
equation
  \be \la{criticalYY} \left[ {\mathbb{D}}+\mathbb{D}^{-1}\right]
  \varphi =- (\mathbb{D}^a+\mathbb{D}^{-a}) \log (1+ e^{- \varphi }).
  \ee
 This equation determines the pseudoenergy $\varphi $ uo to a periodic
 function in $u\to u+ i $.  The ambiguity is lifted by the condition
 that $\varphi \to \vp^\circ$ at infinity.  With this condition the
 saddle point equation gives TBA equation
  \be \la{TBASG} \varphi (u) = RE(u) - \int {dv\over 2\pi} \log(1+
  e^{- \varphi (v)})\, K(v, u).  \ee

Sometimes functional equations as \re{criticalYY} offer a more
effective tool for accessing the solution than the integral TBA
equation.  The method of the ``Quantum Spectral Curve''
\cite{Gromov:2013pga, Gromov:2014caa}, based on a set of functional
equations for Baxter's $Q$-functions, proved to be very successful in
the AdS/CFT integrability.  The $Q$-function, introduced for the
sinh-Gordon model in \cite{Zamolodchikov:2000kt} and
\cite{Lukyanov:2000jp}, is related to the critical value of the
bosonic field by $\vp =- ({\ID^a+\ID^{-a}})\log Q$.  The discrete
Liouville equation implies for the following functional identity for
the $Q$-function,
  \be
 Q  ^{\ID+\ID^{-1}}=
  1+ Q    ^{\ID^{a}+\ID^{-a}},
  \ee
 which can be formulated as a bilinear equation
\be \la{Qsys} Q (u+ {i/2})Q (u- {i/ 2})-Q \left(u+ {i } {a /2}\right)Q
\left(u - {i } {a/ 2}\right) =1.  \ee
 Al.  Zamolodchikov \cite{Zamolodchikov:2000kt} showed using these
 functional equations that $Q (u)$ and $Y(u)= e^{\vp(u)}$ are entire
 functions of $u$ and that the $Q $-system \re{Qsys} implies
 periodicity for the T-functions defined as
 \be
 T(u) = Q^{-1} {(\ID^{1-a}+ \ID ^{ a-1})Q       } ,
 \qquad  \tilde T(u) = Q^{-1} {(\ID^{1+a}+ \ID ^{ a+1})Q       } 
 \ee
   namely $(\ID^a - \ID^{-1})\, T=0$ and $ (\ID^a - \ID )\, \tilde T=0$.

\section{Concluding remarks}
\la{section:conclusions}

In this paper we gave the Thermodynamical Bethe Ansatz a more familiar
to particle physicists formulation based on an effective QFT defined
in the rapidity plane (or Riemann surface).  The elementary
excitations of the effective QFT correspond to particles wrapping the
two main cycles of the torus.  The connected vacuum Feynman diagrams
for the effective QFT are in one-to-one correspondence with the graphs
in the exact cluster expansion for the free energy worked out in
\cite{Kostov:2018ckg, Kostov:2018dmi}.

An intriguing feature of the exact operator representation of the sum
over mirror wrapping particles, eqs.
\re{firstoperrep}-\re{defhatomega}, is that the two logarithmic
factors in the integrand have similar form, although they were derived
by completely different arguments.  Integrating by parts and shifting
the integration variable as $u\to u^{-\g}$ gives the original
integrand with $\bA $ and $\bB $ exchanged.  This remarkable fact
means that if we take the opposite limit, $R$ large and $L$ finite,
and write a similar free-field representation, it will have the same
form as \re{firstoperrep} up to the choice of the integration contour,
which this time should encircle the singularities of the factor $\log
(1+\bB )$.  It is not excluded that the representation
\re{firstoperrep} can be used as a starting point to attack the
problem with finite $R$.

We restricted ourselves to the simplest possible scattering theory
with only one neutral particle.  This restriction is not very
important and there are no conceptual difficulties in generalising the
construction to more interesting cases of non-diagonal scattering and
bound states in the physical/mirror channels, once the Bethe-Yang
equations are diagonalised by the Nested Bethe Ansatz
\cite{Kulish:1983rd}.  There is however a subtle point which is not
yet well understood.  The diagonalisation involves auxiliary particles
of magnonic type with zero momentum and energy.  The auxiliary magnons
should be treated in exactly the same way as the  momentum-carrying
particles, although this is justified only if the number of the  
momentum-carrying particles is large.  Nevertheless this prescription
seems to reproduce correctly the L\"uscher corrections for periodic
\cite{Ahn:2011xq} and even for open \cite{Kostov:2019fvw} boundary
conditions.  Apart of this subtlety the generalisation is
straightforward.  A pair of wrapping operators will be associated with
each node of the Dynkin graph.  The partition function (possibly with
excited states in the physical channel) can be reformulated in terms
of a path integral, which localises to a critical point determined by
the TBA equations.
 
The effective QFT description might be useful in computing a class of
correlation functions in the AdS/CFT integrability.  For example, the
``simplest'' four-point function studied in
\cite{Coronado:2018cxj,Coronado:2018ypq, Belitsky:2019fan,
Bargheer:2019kxb,Bargheer:2019exp} can be formulated as an effective
CFT of real fermions living on the rapidity Riemann surface
\cite{Kostov:2019auq, Kostov:2019stn}.  In this simplest case the
world sheet splits into two or more disks (octagons).

 In general, the world sheet for a correlation function of $n$
 single-trace operators has the topology of a sphere with $n$
 punctures.  Such world sheet hosts a negative curvature $(2-n) \pi$
 distributed among $n-2$ curvature defects represented by the so
 called hexagon operators \cite{BKV1}.  In this case the $\bA$ and the
 $\bB$ operators of the effective QFT will be associated with a system
 of $A$-cycles and $B$-cycles of the world sheet, the $A$-cycles
 connecting pairs of punctures and the $B$-cycles connecting pairs of
 hexagons.  
 
 In \cite{BKV1} and in a series of subsequent works the
 correlation function is expressed as an integral over the mirror
 particles wrapping the $B$-cycles.  The integral is divergent as a
 consequence of the infinite length of the $A$-cycles and requires 
 regularisation.  For a single
 pair of mirror particles, this divergency has been taken under
 control by a point-splitting regularisation of the integral over the
 rapidities \cite{BGK}, but it is not clear how to proceed in the case
 of several mirror particles.  Assuming that the effective CFT
 description  can be generalised to this case, the on-shell condition
 associated with each (asymptotically large) $A$-cycle would give an
 unambiguous prescription about how to regularise the contribution of
 arbitrary number of mirror magnons.

\bigskip

\medskip

\noindent {\bf Acknowledgement.} \ The author thanks Emil Nissimov and
Svetlana Pacheva for useful comments and Zoltan Bajnok, Shota Komatsu,
Didina Serban and Valentina Petkova for important critical remarks on
the manuscript.
 
\appendix

\section{Conventions about the scattering in physical and mirror
kinematics} \la{app:A}

We start with the infinite-volume description where the dispersion
relation between the momentum and the energy of the particles in is
parametrised by an intrinsically complex-valued rapidity variable $u$,
 \be p=p(u), \ E= E(u)\, , \la{disprel} \ee
with $p(-u) = - p(u)$ and $E(-u)= E(u)$.  The mirror transformation
exchanging the space and the time is obtained as an analytic
continuation in the rapidity along some path $\g$.  We adopt the
notations of \cite{BGK} and denote by $u^\g$ the endpoint of the path
which can live in a different sheet of the rapidity Riemann surface.
For a pedagogical explanation see \cite{Komatsu-LesHouches}.  For a
relativistic field theory one can set
  \be p= m\sinh (\pi u) , \quad E= m\cosh (\pi u), \quad u^{ n \g}=
  u+i\, n/2 \qquad \text{(relativistic QFT)}.  \ee
  The momentum and the energy of the mirror theory are given by
 \be \la{mirrorsym} \tilde p(u) = -i E(u^\g), \ \ \ \tilde E(u) = - i
 p(u^\g) \ee
and should take real values for $u$ real.  The square of the mirror
transformation $\g$ is the crossing transformation $\g^2$ exchanging
particles and anti-particles.  The two-particle scattering matrix
$S(u,v)$ is a unimodular function, analytic in the physical strip  
 and satisfying the conditions
 \be\begin{aligned} \la{unitarityS} S(u,v)\, S(v,u )&=1\qquad
 &\text{(unitarity),} \\
 S(u, v^{-\g}) \, S( u, v^{\g}) &= 1 \qquad &\text{(crossing
 unitarity)}.
 \end{aligned}
    \ee
We do not suppose that $S(u,v)$ is a function of the difference of the
rapidities.  We also assume that
\be \la{fermionicproperty} S(u,u) =-1 , \ee
since this is the case with all known scattering theories excepr the
free bosons.  The S-matrix in the mirror channel
  \be \la{mirrorS} \tilde S(u,v) = S(u^\g, v^\g) \ee
  shares the same properties.

\section{Operator representation of the Gaudin determinant}
\la{appendix:C}
 
   Here we will  give the operator representation of
     the determinant of any matrix  of the form
 \be \la{M-matrixB} G_{ik} = G_i \, \d_{ik} - K^- _{ki}, \qquad G_i
 \equiv D_i + \sum_{k=1}^m K^+_{ik} . \ee

 The determinant of the matrix \re{M-matrixB} can be represented as an
 expectation value of a product of $m$ operators composed of the
 bosonic and fermionic gaussian variables, which we denote
 respectively by $\bar\vp_j, \vp_j$ and $\bar\psi_j, \psi_j$, with
 $j=1,..., m$.  If the non-zero bosonic and fermionic correlations are
 given by
\be\begin{aligned} \langle \bar\vp_j\rangle = D_j, \quad \langle \bar
\varphi _j\varphi _k\rangle = K^+_{jk}, \quad \langle \psi_j
\bar\psi_k\rangle = K^-_{jk}, \la{QFTprops}\end{aligned} \ee
then\footnote{A detailed derivation is presented in
\cite{Kostov:2018dmi} where it was used to give a compact proof of the
matrix-tree theorem.}

\be\begin{aligned} \det G &= \left\langle \prod_{j=1}^m ( \bar\varphi
_j -\psi_j \bar\psi_j )\, e^{ \varphi _j} \right\rangle .  \la{QFTrep}
\end{aligned}
 \ee

\medskip

The fermionic and the bosonic variables will be considered as
specialisations of the bosonic and fermionic fields whose correlation
functions given by the scattering phases \be \< \bbvp(u) \bvp(v)\> =
-\tilde \phi(u,v) , \quad \<\bbpsi(u)\bpsi(v)\> = \tilde \phi(u,v) .
\ee
 Namely
 \be \vp_j = -\bvp (u_j)+ \vp^\circ(u_j),\ \bar\vp_j = \bbvp'(u_j) , \
 \psi_j = \bpsi (u_j),\ \bar\psi_j = \bbpsi'(u_j), \ee
and for the correlators
\be K^\pm _{jk} = K^\pm(u_j, u_k) , \ee
where
 \be\begin{aligned} \tilde K^+ (u,v) & \equiv - \< \bbvp'(u) \bvp(v)\>
 = \p_u \tilde \phi(u,v) , \\
  \tilde K^-(u,v) & \equiv \<\bpsi'(u)\bbpsi(v)\> = - \p_u \tilde
  \phi(v,u) \end{aligned}.  \ee
  Then
   \re{QFTrep}   yields
    \begin{align}
   \det G(\uu^\rr)\ \prod _{k=1}^m e^{ -r_k \vp^\circ(u_k)}=
   \left\langle \prod_{k=1}^m \[ \bbvp'(u_j) - r_j \, \bbpsi(u_j)
   \bpsi'(u_j) \] e^{-r_k \bvp (u_k) }\right\rangle .  \la{QFTrepaL}
 \end{align}
This is the general formula in which one should specify the form of
the scattering phases.  For periodic boundary conditions
 \be\begin{aligned} \tilde \phi(u,v) = {1\over i} \log \tilde S(u,v),
 \ \ \tilde K^+ (u,v) = \tilde K^- (u,v) = \tilde K(u,v), \\
  \< \bbvp'(u) \bvp(v)\> = - \tilde K(u,v) , 
  \quad
 \<\bpsi'(u)\bbpsi(v)\> =  \tilde K(u,v).
\end{aligned}
 \ee

 \footnotesize
 

\begin{thebibliography}{10}

\bibitem{Zamolodchikov:1989cf}
A.~B. Zamolodchikov, ``{Thermodynamic Bethe Ansatz in relativistic models.
  Scaling three state Potts and Lee-Yang models},'' {\em Nucl. Phys.} {\bf
  B342} (1990)
695--720.

\bibitem{Zamolodchikov:1991vx}
A.~B. Zamolodchikov, ``{From tricritical Ising to critical Ising by
  thermodynamic Bethe ansatz},'' {\em Nucl. Phys.} {\bf B358} (1991)
524--546.

\bibitem{Zamolodchikov:1991vg}
A.~B. Zamolodchikov, ``{TBA equations for integrable perturbed $SU(2)_k \times
  SU(2)_l / SU(2)_{k+ l}$ coset models},'' {\em Nucl. Phys.} {\bf B366} (1991)
122--134.

\bibitem{YY}
C.~Yang and C.~Yang, ``Thermodynamics of a one-dimensional system of bosons
  with repulsive delta-function interaction,'' {\em Journ. Math. Phys.} {\bf
  10} (1969) 1115.

\bibitem{Integrability-overview-2012}
N.~Beisert {\em et al.}, ``{Review of AdS/CFT Integrability: An Overview},''
  {\em Lett. Math. Phys.} {\bf 99} (2012) 3--32,
\href{http://arXiv.org/abs/hep-th/1012.3982}{{\tt hep-th/1012.3982}}.

\bibitem{Luscher:1985dn}
M.~Luscher, ``{Volume Dependence of the Energy Spectrum in Massive Quantum
  Field Theories. 1. Stable Particle States},'' {\em Commun. Math. Phys.} {\bf
  104} (1986)
177.

\bibitem{2004JPSJ...73.1171K}
G.~{Kato} and M.~{Wadati}, ``{Bethe Ansatz Cluster Expansion Method for Quantum
  Integrable Particle Systems},'' {\em Journal of the Physical Society of
  Japan} {\bf 73} (May, 2004) 1171.

\bibitem{2002JMP....43.5060K}
G.~{Kato} and M.~{Wadati}, ``{Direct calculation of thermodynamic quantities
  for the Heisenberg model},'' {\em Journal of Mathematical Physics} {\bf 43}
  (Oct., 2002) 5060--5078, \href{http://arXiv.org/abs/cond-mat/0212325}{{\tt
  cond-mat/0212325}}.

\bibitem{2001JMP....42.4883K}
G.~{Kato} and M.~{Wadati}, ``{Graphical representation of the partition
  function of a one-dimensional {$\delta$}-function Bose gas},'' {\em Journal
  of Mathematical Physics} {\bf 42} (Oct., 2001) 4883--4893,
  \href{http://arXiv.org/abs/cond-mat/0212323}{{\tt cond-mat/0212323}}.

\bibitem{2001PhRvE..63c6106K}
G.~{Kato} and M.~{Wadati}, ``{Partition function for a one-dimensional
  {$\delta$}-function Bose gas},'' {\em Phys. Rev. E} {\bf 63} (Mar., 2001)
  036106, \href{http://arXiv.org/abs/cond-mat/0212321}{{\tt cond-mat/0212321}}.

\bibitem{Kostov:2018ckg}
I.~Kostov, D.~Serban, and D.-L. Vu, ``{TBA and tree expansion},''
\newblock 2018.
\newblock
\href{http://arXiv.org/abs/arXiv[hep-th]1805.02591}{{\tt
  arXiv[hep-th]1805.02591}}.
\newblock

\bibitem{Kostov:2018dmi}
I.~Kostov, D.~Serban, and D.-L. Vu, ``{Boundary TBA, trees and loops},''
\href{http://arXiv.org/abs/1809.05705}{{\tt 1809.05705}}.

\bibitem{Vu:2018iwv}
D.-L. Vu and T.~Yoshimura, ``{Equations of state in generalized
  hydrodynamics},''
\href{http://arXiv.org/abs/1809.03197}{{\tt 1809.03197}}.

\bibitem{vu2019cumulants}
D.-L. Vu, ``Cumulants of conserved charges in GGE and cumulants of total
  transport in GHD: exact summation of matrix elements?,'' 2019.

\bibitem{Woynarovich:2004gc}
F.~Woynarovich, ``{O(1) contribution of saddle point fluctuations to the free
  energy of Bethe Ansatz systems},'' {\em Nucl. Phys.} {\bf B700} (2004)
  331--360,
\href{http://arXiv.org/abs/cond-mat/0402129}{{\tt cond-mat/0402129}}.

\bibitem{Pozsgay:2010tv}
B.~Pozsgay, ``{On O(1) contributions to the free energy in Bethe Ansatz
  systems: The Exact g-function},'' {\em JHEP} {\bf 08} (2010) 090,
\href{http://arXiv.org/abs/1003.5542}{{\tt 1003.5542}}.

\bibitem{Jiang:2019xdz}
Y.~Jiang, S.~Komatsu, and E.~Vescovi, ``{Structure Constants in $\mathcal{N}=4$
  SYM at Finite Coupling as Worldsheet $g$-Function},''
\href{http://arXiv.org/abs/1906.07733}{{\tt 1906.07733}}.

\bibitem{Nekrasov:2009aa}
N.~A. Nekrasov and S.~L. Shatashvili, ``Quantization of Integrable Systems and
  Four Dimensional Gauge Theories,'' \href{http://arXiv.org/abs/0908.4052}{{\tt
  0908.4052}}.

\bibitem{Balog-TBA}
J.~Balog, ``Field theoretical derivation of the TBA integral equation,'' {\em
  Nuclear Physics B} {\bf 419} (5, 1994) 480--506.

\bibitem{Polyakov:1983tt}
A.~M. Polyakov and P.~B. Wiegmann, ``{Theory of nonabelian Goldstone bosons in
  two dimensions},'' {\em Phys. Lett.} {\bf B131} (1983)
121--126.

\bibitem{Ogievetsky:1984pv}
E.~Ogievetsky, N.~Reshetikhin, and P.~Wiegmann, ``{THE PRINCIPAL CHIRAL FIELD
  IN TWO-DIMENSION AND CLASSICAL LIE ALGEBRA},''. NORDITA-84/38.

\bibitem{Faddeev:1985qu}
L.~D. Faddeev and N.~Y. Reshetikhin, ``{INTEGRABILITY OF THE PRINCIPAL CHIRAL
  FIELD MODEL IN (1+1)-DIMENSION},'' {\em Ann. Phys.} {\bf 167} (1986)
227.

\bibitem{Destri:1994bv}
C.~Destri and H.~J. De~Vega, ``{Unified approach to thermodynamic Bethe Ansatz
  and finite size corrections for lattice models and field theories},'' {\em
  Nucl. Phys.} {\bf B438} (1995) 413--454,
\href{http://arXiv.org/abs/hep-th/9407117}{{\tt hep-th/9407117}}.

\bibitem{Destri:1997yz}
C.~Destri and H.~J. de~Vega, ``{Non-linear integral equation and excited-states
  scaling functions in the sine-Gordon model},'' {\em Nucl. Phys.} {\bf B504}
  (1997) 621--664,
\href{http://arXiv.org/abs/hep-th/9701107}{{\tt hep-th/9701107}}.

\bibitem{paulzinn-thesis}
P.~Zinn-Justin, ``Quelques applications de l'ansatz de Bethe.'' Thesis.

\bibitem{Volin:2010aa}
D.~Volin, ``Quantum integrability and functional equations,''
  \href{http://arXiv.org/abs/1003.4725}{{\tt 1003.4725}}.

\bibitem{Teschner:2008ab}
J.~Teschner, ``On the spectrum of the Sinh-Gordon model in finite volume,''
  {\em Nucl.Phys.B} {\bf 799} (2008) 403--429,
  \href{http://arXiv.org/abs/hep-th/0702214}{{\tt hep-th/0702214}}.

\bibitem{Lukyanov:1993pn}
S.~L. Lukyanov, ``{Free field representation for massive integrable models},''
  {\em Commun. Math. Phys.} {\bf 167} (1995) 183--226,
\href{http://arXiv.org/abs/hep-th/9307196}{{\tt hep-th/9307196}}.

\bibitem{Woynarovich:2010wt}
F.~Woynarovich, ``{On the normalization of the partition function of Bethe
  Ansatz systems},'' {\em Nucl. Phys.} {\bf B852} (2011) 269--286,
\href{http://arXiv.org/abs/1007.1148}{{\tt 1007.1148}}.

\bibitem{Jiang:2019zig}
Y.~Jiang, S.~Komatsu, and E.~Vescovi, ``{Exact Three-Point Functions of
  Determinant Operators in Planar $N=4$ Supersymmetric Yang-Mills Theory},''
\href{http://arXiv.org/abs/1907.11242}{{\tt 1907.11242}}.

\bibitem{Klassen:1989ui}
T.~R. Klassen and E.~Melzer, ``{Purely Elastic Scattering Theories and their
  Ultraviolet Limits},'' {\em Nucl.Phys.} {\bf B338} (1990)
485--528.

\bibitem{Pestun:2016qko}
V.~Pestun, ``{Review of localization in geometry},'' {\em J. Phys.} {\bf A50}
  (2017), no.~44, 443002,
\href{http://arXiv.org/abs/1608.02954}{{\tt 1608.02954}}.

\bibitem{Kostov:2019fvw}
I.~Kostov, D.~Serban, and D.-L. Vu, ``{Boundary entropy of integrable perturbed
  $SU(2)_k$ WZNW},''
\href{http://arXiv.org/abs/1906.01909}{{\tt 1906.01909}}.

\bibitem{Zamolodchikov:2000kt}
A.~B. Zamolodchikov, ``{On the thermodynamic Bethe ansatz equation in
  sinh-Gordon model},'' {\em J. Phys.} {\bf A39} (2006) 12863--12887,
\href{http://arXiv.org/abs/hep-th/0005181}{{\tt hep-th/0005181}}.

\bibitem{Lukyanov:2000jp}
S.~L. Lukyanov, ``{Finite temperature expectation values of local fields in the
  sinh-Gordon model},'' {\em Nucl. Phys.} {\bf B612} (2001) 391--412,
\href{http://arXiv.org/abs/hep-th/0005027}{{\tt hep-th/0005027}}.

\bibitem{Bytsko:2006aa}
A.~Bytsko and J.~Teschner, ``Quantization of models with non-compact quantum
  group symmetry. Modular XXZ magnet and lattice sinh-Gordon model,'' {\em
  J.Phys.A39:12927-12981,2006; J.Phys.A} {\bf 39} (2006) 12927--12982,
  \href{http://arXiv.org/abs/hep-th/0602093}{{\tt hep-th/0602093}}.

\bibitem{Negro:2013wga}
S.~Negro and F.~Smirnov, ``{On one-point functions for sinh-Gordon model at
  finite temperature},'' {\em Nucl.Phys.} {\bf B875} (2013) 166--185,
\href{http://arXiv.org/abs/1306.1476}{{\tt 1306.1476}}.

\bibitem{Bajnok:2019yik}
Z.~Bajnok and F.~Smirnov, ``{Diagonal finite volume matrix elements in the
  sinh-Gordon model},''
\href{http://arXiv.org/abs/1903.06990}{{\tt 1903.06990}}.

\bibitem{Gromov:2013pga}
N.~Gromov, V.~Kazakov, S.~Leurent, and D.~Volin, ``{Quantum Spectral Curve for
  Planar $\mathcal{N} = 4$ Super-Yang-Mills Theory},'' {\em Phys. Rev. Lett.}
  {\bf 112} (2014), no.~1, 011602,
\href{http://arXiv.org/abs/hep-th/1305.1939}{{\tt hep-th/1305.1939}}.

\bibitem{Gromov:2014caa}
N.~Gromov, V.~Kazakov, S.~Leurent, and D.~Volin, ``{Quantum spectral curve for
  arbitrary state/operator in AdS$_{5}$/CFT$_{4}$},'' {\em JHEP} {\bf 09}
  (2015) 187,
\href{http://arXiv.org/abs/hep-th/1405.4857}{{\tt hep-th/1405.4857}}.

\bibitem{Kulish:1983rd}
P.~P. Kulish and N.~Y. Reshetikhin, ``{DIAGONALIZATION OF GL(N) INVARIANT
  TRANSFER MATRICES AND QUANTUM N WAVE SYSTEM (LEE MODEL)},'' {\em J. Phys.}
  {\bf A16} (1983)
L591--L596.

\bibitem{Ahn:2011xq}
C.~Ahn, Z.~Bajnok, D.~Bombardelli, and R.~I. Nepomechie, ``{TBA, NLO Luscher
  correction, and double wrapping in twisted AdS/CFT},'' {\em JHEP} {\bf 12}
  (2011) 059,
\href{http://arXiv.org/abs/1108.4914}{{\tt 1108.4914}}.

\bibitem{Coronado:2018cxj}
F.~Coronado, ``{Bootstrapping the simplest correlator in planar $\mathcal N =
  4$ SYM at all loops},''
\href{http://arXiv.org/abs/hep-th/1811.03282}{{\tt hep-th/1811.03282}}.

\bibitem{Coronado:2018ypq}
F.~Coronado, ``{Perturbative four-point functions in planar $ \mathcal{N}=4 $
  SYM from hexagonalization},'' {\em JHEP} {\bf 01} (2019) 056,
\href{http://arXiv.org/abs/hep-th/1811.00467}{{\tt hep-th/1811.00467}}.

\bibitem{Belitsky:2019fan}
A.~V. Belitsky and G.~P. Korchemsky, ``{Exact null octagon},''
\href{http://arXiv.org/abs/1907.13131}{{\tt 1907.13131}}.

\bibitem{Bargheer:2019kxb}
T.~Bargheer, F.~Coronado, and P.~Vieira, ``{Octagons I: Combinatorics and
  Non-Planar Resummations},''
\href{http://arXiv.org/abs/1904.00965}{{\tt 1904.00965}}.

\bibitem{Bargheer:2019exp}
T.~Bargheer, F.~Coronado, and P.~Vieira, ``{Octagons II: Strong Coupling},''
\href{http://arXiv.org/abs/1909.04077}{{\tt 1909.04077}}.

\bibitem{Kostov:2019auq}
I.~Kostov, V.~B. Petkova, and D.~Serban, ``{The Octagon as a Determinant},''
\href{http://arXiv.org/abs/1905.11467}{{\tt 1905.11467}}.

\bibitem{Kostov:2019stn}
I.~Kostov, V.~B. Petkova, and D.~Serban, ``{Determinant formula for the octagon
  form factor in $\mathcal{N}=4$ SYM},''
\href{http://arXiv.org/abs/1903.05038}{{\tt 1903.05038}}.

\bibitem{BKV1}
B.~Basso, S.~Komatsu, and P.~Vieira, ``{Structure Constants and Integrable
  Bootstrap in Planar N=4 SYM Theory},''
\href{http://arXiv.org/abs/1505.06745}{{\tt 1505.06745}}.

\bibitem{BGK}
B.~Basso, V.~Goncalves, and S.~Komatsu, ``{Structure constants at wrapping
  order},''
\href{http://arXiv.org/abs/1702.02154}{{\tt 1702.02154}}.

\bibitem{Komatsu-LesHouches}
S.~Komatsu, ``{Lectures on Three-point Functions in N=4 Supersymmetric
  Yang-Mills Theory},''
\href{http://arXiv.org/abs/1710.03853}{{\tt 1710.03853}}.

\end{thebibliography}
%

\providecommand{\href}[2]{#2}\begingroup\raggedright\endgroup

\end{document}